\documentclass{llncs}

\usepackage{amssymb}

\usepackage[cmex10]{amsmath}
\usepackage{array}
\usepackage{url}
\usepackage{cite}
\usepackage{color}
\usepackage{proof}
\usepackage{graphicx}
\usepackage{xspace}
\usepackage{subfigure}
\usepackage{mycommands}

\title{(Mathematical) Logic for Systems Biology \\ (Invited Paper)
}

\author{Jo{\"e}lle Despeyroux
   \thanks{CMSB 2016, 
   Cambridge, UK. LNBI 9859. DOI 10.1007/978-3-319-45177-0}
}
\institute{INRIA and CNRS, I3S, Sophia-Antipolis, France \\ joelle.despeyroux@inria.fr}

\authorrunning{Despeyroux}
        
\begin{document}
\maketitle

\begin{abstract}
We advocates here the use of (mathematical) logic for systems biology, 
as a unified framework well suited for both modeling the dynamic behaviour of 
biological systems, expressing properties of them, and verifying these 
properties. 
The potential candidate logics should have a traditional proof theoretic 
pedigree (including a sequent calculus presentation enjoying cut-elimination 
and focusing), and should come with (certified) proof tools. Beyond providing 
a reliable framework, this allows the adequate encodings of our biological 
systems.
We present 
two candidate logics (two modal extensions of linear logic,
 called HyLL and SELL), along with biological examples. 
The examples we have considered so far are very simple ones 
- coming with completely formal (interactive) proofs in Coq.
Future works includes using automatic provers, which would extend existing 
automatic provers for linear logic. This should enable us to specify and study 
more realistic examples in systems biology, biomedicine (diagnosis and 
prognosis), and eventually neuroscience.
\end{abstract}


\section{Introduction} \label{sec:intro}
We consider here the question of reasoning about 
biological systems in (mathematical) logic.
We show that two new logics, both modal extensions of 
linear logic \cite{girard87tcs} (LL),
are particularly well-suited to this purpose.
The first logic, called Hybrid Linear Logic (HyLL), has been developed by the 
author in joint work with K. Chaudhuri \cite{ChaudhuriDespeyroux:14}.
The second logic, an extension of Subexponential Linear Logic ($\sellU$),
has been independently proposed by C. Olarte, E. Pimentel and 
V. Nigam \cite{NigamOlartePimentel:concur-13}. 
Both HyLL and SELL provide a unified
framework to encode biological systems, to express temporal properties
of their dynamic behaviour, and to prove these properties.  
By constructing proofs in the logics, 
we directly witness reachability as logical entailment
\cite{deMaria-Despeyroux-Felty:14-fmmb,DBLP:journals/entcs/ChiarugiFHO16}.  
This approach is in contrast to
most current approaches to applying formal methods to systems biology,
which generally 
encode biological systems either in a 
dedicated programming language
\cite{PC05tcsb,FS08sfm,Danos:09-tcs},
or in differential equations \cite{CampagnaPiazza:entcs09},
express properties in a temporal logic, 
and then verify these properties against some form of traces 
(model-checking), 
eventually built using an external simulator.

In a joint work with E. De Maria and A. Felty,
we presented some first applications of HyLL to systems biology 
\cite{deMaria-Despeyroux-Felty:14-fmmb}.  
In these first experiments, we focused on Boolean systems and 
in this case a time unit corresponds to a transition in the system.
We believe that discrete modeling is crucial in systems
biology because it allows taking into account some phenomena
that have a very low chance of happening (and could thus be neglected by
differential approaches), but which may have a strong impact on system
behavior.

In a recent joint work with C. Olarte and E. Pimentel
\cite{Despeyroux-Olarte-Pimentel:lsfa-16},
we compared HyLL and SELL, providing two encodings. 
The first enoding is from HyLL's logical rules into LL with the highest level 
of adequacy, hence showing that HyLL is as expressive as LL.
We also proposed an encoding of HyLL into \sellU\ ,
showing 
that \sellU\ is more expressive than HyLL. 
However, the simplicity of HyLL might be of interest, 
both from the user point of view and
as far as proof search is concerned 
(a priori easier and more efficient in HyLL than in SELL).
In this joint work, we 
furthermore encoded temporal operators of Computational Tree Logic (CTL)  
into linear logic with fixed point operators.

We first recall here these two previous works.
Then we briefly mention our current joint work with P. Lio, 
on formalizing the evolution of cancer cells,
concluding with some future work.

This note is thus based on joint works with 
K. Chaudhuri (INRIA Saclay), A. Felty (Univ. of Ottawa), 
P. Lio (Cambridge Univ.), and C. Olarte and 
E. Pimentel (Universidade Federal do Rio Grande do Norte, Brazil).

\section{Preliminaries} 

Although we assume that the reader is familiar with linear logic
\cite{girard87tcs} (LL), we review some of its basic proof theory in the 
following sections.
First, let us gently introduce linear logic by means of an example.

\subsection{Linear Logic for Biology}
\label{subsec:ll}

Linear Logic (LL) \cite{girard87tcs} is particularly well suited for describing 
state transition systems. 
LL  has been successfully used to model such diverse systems as: 
the $\pi$-calculus, 
concurrent ML, 
security protocols, 
multiset rewriting, 
and games.

In the area of biology, 
a rule of activation 
(e.g., a protein activates a gene or the transcription of another protein)
can be modeled by the following LL axiom:
$$\cn{active}(a,b) \eqdef \cn{pres}(a) \limp (\cn{pres}(a) \otimes \cn{pres}(b)).$$
The formula $\cn{active}(a,b)$ describes the fact that a state where 
$a$ is present ($\cn{pres}(a)$ is true)
can evolve into a state where both  $\cn{pres}(a)$ and $\cn{pres}(b)$ are 
true.

Propositions such as $\cn{pres}(a)$ are called {\it resources}, and a rule in the logic 
can be viewed as a rewrite rule from a set of resources into another set of resources,
where a set of resources 
describes a state of the system.
Thus, a particular state transition system can be modeled by a set of rules of 
the above shape. The rules of the logic then allow us to prove some 
desired properties of the system, such as, for example, the existence of a stable state.
However, linear implication is timeless. 
Linear implication can be used to model one event
occurring after another, but it cannot be precise about how many steps
or how long the delay is without explicitly encoding time.
In a domain where
resources have lifetimes and state changes have temporal, probabilistic or
stochastic \emph{constraints}, then the logic will allow inferences that may not
be realizable in the system being modeled. 
This was the motivation of 
the development of HyLL, which was designed
to represent constrained transition systems.

\subsection{Linear Logic and Focusing} \label{sec:ll}
\emph{Literals} are either atomic formulas ($p$) or their
negations ($p^\bot$).  The connectives $\tensor$ and $\lpar$ and their units $1$
and $\bot$ are \emph{multiplicative}; the connectives $\plus$ and
$\with$ and their units $0$ and $\top$ are \emph{additive}; 
$\forall$ and $\exists$ are (first-order) quantifiers;
and $\bang$ and $\quest$ are the exponentials (called bang and question-mark,
respectively).  

First proposed by Andreoli \cite{DBLP:journals/logcom/Andreoli92} for linear logic,
focused proof systems provide  normal form proofs for cut-free proofs.
The connectives of linear logic can be divided into two classes.  The
{\em negative} connectives have invertible introduction rules: these
connectives are $\lpar$, $\bottom$, $\with$, $\top$, $\forall$, and
$\quest$.  The {\em positive} connectives 
$\otimes$, $\one$, $\oplus$,
$\zero$, $\exists$, and $\bang$
are the de Morgan duals of
the negative connectives.
  A formula is {\em positive} if it is
a negated atom or its top-level logical connective is positive.
Similarly, a formula is {\em negative} if it is an atom or its
top-level logical connective is negative. 

Focused proofs are organized into two \emph{phases}.  In the \emph{negative} phase, all the invertible inference rules are eagerly applied. 
The \emph{positive} phase begins by choosing  a positive formula $F$ on which to focus. Positive rules are applied to $F$ until either $\one$ or a negated atom is encountered (and the proof must end by applying the initial rules),  
the promotion rule ($\bang$) is applied,  or a negative subformula is encountered and the proof switches to the negative phase.

This change of phases on proof search is particularly interesting when the 
focused formula is a {\em bipole}~\cite{DBLP:journals/logcom/Andreoli92}.
%
%
Focusing on a bipole will produce a
single positive and a single negative phase. This two-phase decomposition 
enables us to  adequately capture  the application of object-level inference 
rules by the meta-level linear logic, 
as shown in \cite{Despeyroux-Olarte-Pimentel:lsfa-16}.



%
\subsection{Hybrid Linear Logic} \label{sec:hyll}
Hybrid Linear Logic  (HyLL)
is a conservative extension of Intuitionistic
first-order Linear Logic (ILL)~\cite{girard87tcs} where the truth judgments 
are labelled by worlds representing constraints on states and state transitions. 
Instead of the ordinary judgment ``$A$ is true'', for a proposition $A$,
judgments of HyLL are of the form 
``$A$ is true at world $w$'', abbreviated as $A ~@~ w$. 
Particular choices of worlds
produce particular instances of HyLL. 
Typical examples  are ``$A$ is true at time $t$'', 
or ``$A$ is true with probability $p$''.
HyLL was first proposed in~\cite{ChaudhuriDespeyroux:14} and it has  been used 
as a logical framework for specifying biological
systems~\cite{deMaria-Despeyroux-Felty:14-fmmb}.

Formally, worlds are defined as follows.
\begin{definition}[HyLL worlds] \label{def:constraint-domain}
  A \emph{constraint domain} $\cal W$ is a monoid structure $\langle W, ., \iota\rangle$. 
  The elements of $W$ are called \emph{worlds}
  and its \emph{reachability relation}  $\preceq\ : W \times W$ is defined as  $u \preceq w$ if there exists 
  $v \in W$ such that $u . v = w$. 
\end{definition}
\noindent
The identity world $\iota$ is $\preceq$-initial and is intended to represent the
lack of any constraints. Thus, the ordinary first-order linear logic is embeddable into any 
instance of HyLL by setting all world labels to the identity.
A typical, simple example of constraint domain is 
$\mathcal{T} = \langle \N, +, 0\rangle$, representing instants of time. 

Atomic propositions $(p,q,\ldots)$ are applied to a sequence of terms 
$(s,t,\ldots)$, which are drawn from an untyped term language containing 
constants  $(c,d,\ldots)$, term variables $(x, y, \ldots)$ and function symbols 
$(f, g, \ldots)$ applied to a list of terms $(\vec{t})$. 
Non-atomic propositions are constructed from the connectives of first-order
intuitionistic linear logic and the two hybrid connectives 
\emph{satisfaction} ($\texttt{at}$), which states that a
proposition is true at a given world ($w, \iota, u.v, \ldots$), and
\emph{localization} ($\downarrow$), which binds a name for the (current) world 
the proposition is true at. 
The following grammar summarizes the syntax of HyLL.

\begin{tabular}{l@{\ }r@{\ }l}
  $t$ & $::=$ & 
       $c ~|~ x ~|~ f(\vec t)$ \\
  $A, B$ & $::=$ & 
       $p(\vec t) ~|~ A \otimes B ~|~ \mathbf{1} ~|~ A \rightarrow B ~|~ 
                A \mathbin{\&} B ~|~ \top ~|~
       A \oplus B ~|~  \mathbf{0} ~|~ ! A ~|~ $  \\
       & &
       $\forall x.~ A ~|~ \exists x.~ A ~|~ $

       $(A ~\at~ w) ~|~ \downarrow u.~ A ~|~ \forall u.~ A ~|~ \exists u.~ A$ \\
\end{tabular}

\noindent
Note that world $u$ is bounded in 
the propositions $\downarrow u.~A$, $\forall u.~ A$ and $\exists u.~ A$.
World variables cannot be used in terms, and neither can term variables occur in 
worlds. This restriction is important for the modular design of HyLL because it 
keeps purely logical truth separate from constraint truth.  
%
We note that $\downarrow$ and $\at$ commute freely with all non-hybrid 
connectives~\cite{ChaudhuriDespeyroux:14}.  


The sequent calculus  \cite{Gentzen35} presentation of HyLL uses sequents of
the form $\Gamma; \Delta \vdashseq C ~@~ w$ where 
$\Gamma$ (\emph{unbounded context}) is a set and 
$\Delta$ (\emph{linear context}) is a multiset
of judgments of the form $A ~@~ w$.
Note that in a judgment $A ~@~ w$ (as in a proposition $A ~\at~ w$), $w$ can be 
any expression in $\cal W$, not only a variable.

The inference rules dealing with the new hybrid connectives are
depicted below 
(the complete set of rules can be found in 
\cite{ChaudhuriDespeyroux:14}).
$$
  \dfrac{\Gamma ; \Delta \vdashseq A @ u} 
        {\Gamma ; \Delta \vdashseq (A ~\at~ u) @ w} \at R
  ~ \qquad
  \dfrac{\Gamma ; \Delta, A @ u \vdashseq C @ w} 
        {\Gamma ; \Delta, (A ~\at~ u) @ v \vdashseq C @ w} \at L
$$
$$
  \dfrac{\Gamma ; \Delta \vdashseq A [w / u] @ w} 
        {\Gamma ; \Delta \vdashseq \downarrow u. A @ w} \downarrow R
  ~  \qquad
  \dfrac{\Gamma ; \Delta, A [v / u] @ v \vdashseq C @ w}
        {\Gamma ; \Delta, \downarrow u. A @ v \vdashseq C @ w} \downarrow L
$$
Note that $(A ~\at~ u)$ is a \emph{mobile} proposition:
it carries with it the world at which it is true. 
%
%
Weakening and contraction are admissible rules for the unbounded context.  

The most important structural properties are the admissibility of 
the general identity (i.e. over any formulas, not only atomic propositions) and cut theorems. 
While the first provides a syntactic completeness theorem for the logic, 
the latter guarantees consistency
(i.e. that there is no proof of  $ . ; . \vdashseq \mathop{0} ~@~ w$).

\begin{theorem}[Identity/Cut] 
  \label{thm:cut} 
  $
   \\
   1.~\Gamma ; A ~@~ w \vdashseq A ~@~ w\\
   2.~ {If}~ \Gamma ; \Delta \vdashseq A ~@~ u 
   ~{and}~ \Gamma ; \Delta', A ~@~ u \vdashseq C ~@~ w, 
   ~{then}~ \Gamma ; \Delta, \Delta' \vdashseq C ~@~ w \\
   3.~ {If}~ \Gamma ; . \vdashseq A ~@~ u 
   ~{and}~ \Gamma, A ~@~ u ; \Delta \vdashseq C ~@~ w, 
   ~{then}~ \Gamma ; \Delta \vdashseq C ~@~ w.
$
\end{theorem}
Moreover,
HyLL is conservative with respect to intuitionistic linear logic: as long as no 
hybrid connectives are used, the proofs in HyLL are identical to those in ILL.
It is worth noting that HyLL 
is more expressive than S5, as it allows direct manipulation of the worlds using
the hybrid connectives and HyLL's $\delta$ connective 
(see Section~\ref{sec:temporal}) is not definable in S5.
We also note that HyLL admits a complete focused  \cite{DBLP:journals/logcom/Andreoli92} proof system.
The interested reader can find proofs and further meta-theoretical theorems 
about HyLL in~\cite{ChaudhuriDespeyroux:14}.

%

\paragraph{Modal Connectives.}

We can define modal connectives in HyLL as follows:
\bgroup 
\begin{definition}[Modal connectives] \label{def:modal-connectives} \mbox{} \\
\vspace{-1ex}
$$
     \Box A \eqdef {\downarrow} u.~ \forall w.~ (A ~\at~ u . w) \quad
     \Diamond A \eqdef {\downarrow} u.~ \exists w.~ (A ~\at~ u . w) \quad
     \delay{v} A \eqdef {\downarrow} u.~ (A ~\at~ u . v)
$$
\end{definition}
\egroup

\noindent
$\Box A$ [resp. $\Diamond A$] represents all [resp. some] state(s) 
satisfying $A$ and reachable from now. 
%
The connective $\delta$ represents a form of delay.
%


%
\subsection{Subexponentials in Linear Logic}
\label{sec:sell} 
Linear logic with subexponentials
(\sell) shares with LL  all 
its connectives except the exponentials:
instead of having a single pair of exponentials $\bang$ and $\quest$, \sell\ may
contain as many \emph{subexponentials}
 \cite{danos93kgc,OlartePimentelNigam:tcs-15}, 
written $\nbang{a}$  and $\nquest{a}$, as one needs.  
The grammar of formulas in  \sell\  is as follows: 
\[\begin{array}{lcl}
 F &::=& \zero \mid \one \mid  \top \mid \bottom \mid p(\vec t) \mid F_1 \tensor F_2 \mid F_1 \oplus F_2 \mid F_1 \lpar F_2 \mid  F_1 \with F_2 \mid \\
& & \exists x. F \mid \forall x. F\mid  
  \nbang{a} F \mid \nquest{a} F 
 \end{array}
\]
The proof system for \sell\ is specified by a
\emph{subexponential signature} $\Sigma = \tup{I, \preceq, U}$, where $I$
is a set of labels, $U \subseteq I$ is a set specifying which
subexponentials allow weakening and contraction, and $\preceq$ is a
pre-order among the elements of $I$. We shall use $a,b,\ldots$
 to range over elements in $I$ and we will assume that $\preceq$
is upwardly closed with respect to $U$, \ie, if $a \in U$ and $a \preceq
b$, then $b \in U$. 

The system $\sell$ is constructed by adding all the rules for
the linear logic connectives  except for the exponentials. 
The rules for subexponentials are 
dereliction and
promotion of the subexponential labelled with $a \in I$
\[
\infer[\nbang{a}]{\vdash\nquest{a_1} F_1, \ldots \nquest{a_n} F_n, \nbang{a} G}{\vdash\nquest{a_1} F_1, \ldots \nquest{a_n} F_n,  G}
\qquad
\infer[\nquest{a}]{\vdash\Gamma, \nquest{a} G}{\vdash\Gamma, G} 
\]
Here, the rule $\nbang{a}$ has the side condition
that $a\preceq {a_i}$ for all $\tsl{i}$. That is, one can only
introduce a $\nbang{a}$ on the right  if
all other formulas in the sequent are marked with indices that are
greater or equal than $a$. Moreover, for all 
indices $a \in U$, we add the usual  rules for weakening and contraction. 

We can enhance  the expressiveness of SELL with the subexponential quantifiers $\forallLoc$ and $\existsLoc$ (\cite{NigamOlartePimentel:concur-13,OlartePimentelNigam:tcs-15})
given by the rules (omitting the subexponential signature)
\[
 \infer[\forallLoc]{\vdash\Gamma, \forallLoc \typeloc{l_x}{a}. G}
 {\vdash \Gamma , G[l_{e}/l_x]}
\qquad
 \infer[\existsLoc]{ \vdash\Gamma, \existsLoc \typeloc{l_x}{a}. G}
 {\vdash\Gamma , G[l/l_x]}
\]
where   $l_e$ is fresh. 
Intuitively, subexponential variables play a similar role as eigenvariables. 
The generic variable $\typeloc{l_x}{a}$ represents any subexponential,
constant or variable in the ideal of $a$. Hence $l_x$ can be substituted 
by any  subexponential $l$ of type $b$ (i.e., $l:b$) if 
$b\preceq a$. 
We call the resulting system $\sellU$.

As shown in \cite{NigamOlartePimentel:concur-13,OlartePimentelNigam:tcs-15}, 
\sellU\ admits a cut-free and also a complete focused proof system.

 \begin{theorem}
$\sellU$ admits cut-elimination for any subexponential signature. 
\end{theorem}

\paragraph{Modal connectives.} 
We can define modal connectives in SELL as follows: 
$$
       \Box_u A  \eqdef \forall l:u.~ \nbang{l}A \quad
       \Diamond_u A  \eqdef \exists l:u.~ \nbang{l}A \quad
       \Box A  \eqdef \forall t:\infty.~ \nbang{t}A \quad
       \Diamond A  \eqdef \exists t:\infty.~ \nbang{t}A
$$ 


\section{First experiments with HyLL}
In a joint work with E. De Maria and A. Felty,
we presented some first applications of HyLL to systems biology 
\cite{deMaria-Despeyroux-Felty:14-fmmb}.  
In these first experiments, we focused on Boolean systems and 
in this case a time unit corresponds to a transition in the system.

The activation rule seen in LL (Sec. \ref{subsec:ll}) can be written in HyLL as
$$\cn{active}(a,b) \eqdef 
\cn{pres}(a) \limp \delay{1}~ (\cn{pres}(a) \otimes \cn{pres}(b)).$$

We chosed a simple yet representative biological example concerning
the DNA-damage repair mechanism based on proteins p53 and Mdm2, 
and present and proved several properties of this system.
All these properties were reachability properties or the existence of an 
invariant. Most interesting proofs require induction or case analysis, 
that we borrowed from the meta-level (Coq).
We fully formalized these proofs 
in the Coq Proof Assistant \cite{BertotCasteran:2004}.
In Coq, we can both reason in HyLL and 
formalize meta-theoretic properties about it.

We discussed the merits
and eventual drawbacks of this new approach compared to approaches using 
temporal logic and model checking.
To better illustrate the correspondence with such approaches, 
which all use temporal logic to reason about 
(simulations of models of) the biological systems described, 
we also 
presented, informally but in some detail,
the encoding of temporal logic operators in HyLL. 

\section{Relative Expressiveness Power of HyLL and SELL} 
\label{sec:hyll-sell}

We observe that,  while linear logic has only seven logically distinct prefixes
of bangs and question-marks, \sell\ allows for an
unbounded number of such prefixes, \eg, $\nbang{i}$, or
$\nbang{i}\nquest{j}$. 
Hence,  by using different prefixes, 
we  allow for the  specification of  richer systems where subexponentials are
used to mark different modalities/states. For instance,
subexponentials can be used to represent contexts of proof
systems~\cite{DBLP:journals/entcs/NigamPR11}; to 
specify systems with temporal, epistemic and spatial modalities 
\cite{OlartePimentelNigam:tcs-15} 
and to specify and verify biological 
systems \cite{DBLP:journals/entcs/ChiarugiFHO16}.
An inhibition rule can be written in (classical) SELL as
$$\cn{inhib}(a,b) \eqdef 
   \nbang{t} a \limp \nbang{t+1} (a \otimes b^\bottom).$$

\paragraph{HyLL and Linear Logic.}
One may wonder whether the use of worlds in HyLL  increases also the 
expressiveness of LL. In a joint work with C. Olarte and E. Pimentel 
\cite{Despeyroux-Olarte-Pimentel:lsfa-16},
we proved that this is not the case, by 
showing that HyLL rules can be directly encoded into LL by using the methods 
proposed in \cite{DBLP:journals/tcs/MillerP13}. 
Moreover, the encoding of HyLL into LL
is {\em adequate} in the sense that a focused step in LL corresponds 
{\em exactly} to the application of one inference rule in HyLL. 

\paragraph{HyLL and SELL.}
Linear logic allows for the specification of two kinds of context maintenance: 
both weakening and contraction are available (classical context)  or 
neither is available (linear context). 
That is,  when we encode (linear) judgments in HyLL belonging to 
different worlds, the resulting meta-level atomic formulas will be stored in 
the same (linear) LL context. The same happens with classical  HyLL judgments  
and the classical LL context.

Although this is perfectly fine, 
encoding  HyLL into \sellU\  allows for a better understanding  of worlds 
in HyLL. 
For that, we use subexponentials to represent worlds, 
having each world as a  linear context. 
A HyLL judgment of the shape $F@w$ in the (left)
linear context is encoded as the \sellU\ formula $\nquest{w}\lF{F@w}$. 
Hence, HyLL judgments that hold at world $w$ are stored at the $w$ 
linear context of \sellU. A judgment of the form $G@w$ in the
classical HyLL context is encoded as the \sellU\ formula 
$\nquest{\copysell}\nquest{w}\lF{G@w}$. Then, the encoding of $G@w$ 
is stored in the unbounded (classical) subexponential context $\copysell$. 

We showed that our encoding is indeed adequate.
Moreover, as before, 
the adequacy of the encodings is on the {\em level of derivations}.

\paragraph{Information Confinement.} 
One of the features needed to specify spatial modalities  is information 
\emph{confinement}: a space/world can be inconsistent and this does not imply 
the inconsistency of the whole system. 
We showed in \cite{Despeyroux-Olarte-Pimentel:lsfa-16}
that information confinement cannot be specified in HyLL. 
The authors in   \cite{NigamOlartePimentel:concur-13} exploit the combination 
of subexponentials of the form $\nbang{w}\nquest{w}$  in order to specify 
information confinement in \sellU. 
  More precisely, note that the sequents (in a 2-sided presentation of \sell) 
$\nbang{w}\nquest{w}\zero \not\vdash \zero$ 
and $\nbang{w}\nquest{w}\zero \not\vdash \nbang{v}\nquest{v}\zero$,  
representing ``inconsistency is local'' and `
`inconsistency is not propagated'' respectively hold in SELL. 


\section{Computation Tree Logic (CTL) in Linear Logic.}
\label{sec:temporal}

Hybrid linear logic is expressive enough to encode some forms of modal 
operators, thus allowing for the specification of properties of transition 
systems. As mentioned in \cite{deMaria-Despeyroux-Felty:14-fmmb}, 
it is possible to encode CTL temporal operators into HyLL considering 
existential ($\tE$) and bounded universal ($\tA$) path quantifiers. 
We extended these encodings
in \cite{Despeyroux-Olarte-Pimentel:lsfa-16}, showing how to fully 
capture $\tE$ and $\tA$ CTL quantifiers in linear logic with fixed points. 
For that, we used the system $\mu$MALL \cite{DBLP:journals/tocl/Baelde12} 
that extends MALL
(multiplicative, additive linear logic) with fixed point operators.
In \cite{deMaria-Despeyroux-Felty:14-fmmb}, 
proofs of (encodings of) properties involving CTL quantifiers 
use induction borrowed from the (Coq) meta-level.
In \cite{Despeyroux-Olarte-Pimentel:lsfa-16}, we could directly use 
fixed points in linear logic.

\section{Concluding Remarks and Future Work} 
\label{sec:conclusion}

Concerning related work, it is worth noticing that there are some other 
logical frameworks that are extensions of LL, for example,  
HLF\cite{reed06hylo}.
Being a logic in the LF family, HLF is based on natural deduction, hence
having a complex notion of ($\beta \eta$) normal forms.
Thus adequacy (of encodings of systems) results are often much harder to 
prove in HLF than in (focused) HyLL/SELL.
%
HLF seems to have been later abandonned in favour of Hybridized Intuitionistic 
Linear Logic (HILL) \cite{Caires-Perez-Pfenning:hill14-sub}
- a type theory based on a subpart of HyLL. 
  
%
Both HyLL and \sell\ have been used for formalizing and analyzing biological systems 
\cite{deMaria-Despeyroux-Felty:14-fmmb,DBLP:journals/entcs/ChiarugiFHO16}. 
\sell\ proved to be a broader framework for handling such systems 
(in particular localities).
However, the simplicity of HyLL may be of interest  for specific purposes, 
such as building tools for diagnosis in biomedicine.
%
%
%

Formal proofs in HyLL were implemented in 
\cite{deMaria-Despeyroux-Felty:14-fmmb}, in the 
Coq~\cite{BertotCasteran:2004} proof assistant.
It would be interesting to extend the implementations of HyLL given there to 
\sell.
Such an interactive proof environment
would enable both formal studies of encoded systems in \sell\ and formal 
meta-theoretical study of \sell\ itself.

We may pursue the goal of using HyLL/\sell\ for further applications. 
That might include neuroscience, a young and promising
science where many hypotheses are provided and need to be verified. 
Indeed, logic is a general tool whose area of potential applications are not 
restricted per se. 
This is in contrast to most of the other approaches, which are valid only 
in a restricted area (typically inside or outside the cell).


In an ongoing joint work with P. Lio, we are formalizing the evolution of 
cancer cells, acquiring driver or passenger mutations. 
A rule describing an intravasating Circulating Tumour Cell, for example,
might be:
$$\cn{C}(n,breast,f,[{\small \texttt{EPCAM}}]) 
   \limp \delay{d} \cn{C}(n,blood,1,[{\small \texttt{EPCAM}}])
$$
where $f$ is a fitness parameter, here in $\{0,1\}$.
Our long term goal here is the design of 
a Logical Framework for disease diagnosis and therapy prognosis.
This requires the development of automatic tools for proof search in our logics.
These tools should benefit both 
from current research on 
proof search in linear logic 
and from current developments of automatic provers for SELL.

\bibliographystyle{splncs03}
\bibliography{references} \vfill

\end{document}